\documentclass[twocolumn,english,aps,prb,showpacs,amsmath,amssymb,superscriptaddress]{revtex4}
\usepackage[latin9]{inputenc}
\usepackage{color}
\usepackage{amstext}
\usepackage{amssymb}
\usepackage{graphicx}

\makeatletter
\@ifundefined{textcolor}{}
{%
 \definecolor{BLACK}{gray}{0}
 \definecolor{WHITE}{gray}{1}
 \definecolor{RED}{rgb}{1,0,0}
 \definecolor{GREEN}{rgb}{0,1,0}
 \definecolor{BLUE}{rgb}{0,0,1}
 \definecolor{CYAN}{cmyk}{1,0,0,0}
 \definecolor{MAGENTA}{cmyk}{0,1,0,0}
 \definecolor{YELLOW}{cmyk}{0,0,1,0}
}

\usepackage{color}

\newcommand{\beq}{\begin{eqnarray}}
\newcommand{\eeq}{\end{eqnarray}}

\makeatother

\usepackage{babel}
\begin{document}

\title{Temperature and polarization dependence of low-energy magnetic fluctuations in nearly-optimal-doped NaFe$_{0.9785}$Co$_{0.0215}$As}

\author{Yu Song}

\email{Yu.Song@rice.edu}

\selectlanguage{english}%

\affiliation{Department of Physics and Astronomy, Rice University, Houston, Texas
77005, USA}

\author{Weiyi Wang}

\affiliation{Department of Physics and Astronomy, Rice University, Houston, Texas
77005, USA}

\author{Chenglin Zhang}

\affiliation{Department of Physics and Astronomy, Rice University, Houston, Texas
77005, USA}

\author{Yanhong Gu}

\affiliation{Beijing National Laboratory for Condensed Matter Physics, Institute of Physics, Chinese Academy of Sciences, Beijing 100190, China}
\affiliation{School of Physical Sciences, University of Chinese Academy	of Sciences, Beijing 100190, China}

\author{Xingye Lu}

\affiliation{Center for Advanced Quantum Studies and Department of Physics, Beijing
	Normal University, Beijing 100875, China}

\author{Guotai Tan}

\affiliation{Center for Advanced Quantum Studies and Department of Physics, Beijing
Normal University, Beijing 100875, China}

\author{Yixi Su}

\affiliation{J{ü}lich Centre for Neutron Science, Forschungszentrum J{ü}lich GmbH, Outstation at MLZ, D-85747 Garching, Germany}

\author{Fr\'{e}d\'{e}ric Bourdarot}

\affiliation{Institut Laue Langevin, 71 Avenue des Martyrs, 38042 Grenoble, France}

\author{A. D. Christianson}

\affiliation{Quantum Condensed Matter Division, Oak Ridge National Laboratory, Oak Ridge, Tennessee 37831, USA}

\author{Shiliang Li}
\affiliation{Beijing National Laboratory for Condensed Matter Physics, Institute of Physics, Chinese Academy of Sciences, Beijing 100190, China}
\affiliation{School of Physical Sciences, University of Chinese Academy	of Sciences, Beijing 100190, China}
\affiliation{Collaborative Innovation Center of Quantum Matter, Beijing, China}

\author{Pengcheng Dai}

\email{pdai@rice.edu}

\selectlanguage{english}%

\affiliation{Department of Physics and Astronomy, Rice University, Houston, Texas
77005, USA}

\affiliation{Center for Advanced Quantum Studies and Department of Physics, Beijing
Normal University, Beijing 100875, China}

\begin{abstract}
	
We use unpolarized and polarized neutron scattering to study the temperature and polarization dependence of low-energy magnetic fluctuations in nearly-optimal-doped NaFe$_{0.9785}$Co$_{0.0215}$As, with coexisting superconductivity ($T_{\rm c}\approx19$ K) and weak antiferromagnetic order ($T_{\rm N}\approx30$ K, ordered moment $\approx0.02$ $\mu_{\rm B}$/Fe). A single spin resonance mode with intensity tracking the superconducting order parameter is observed, although energy of the mode only softens slightly on approaching $T_{\rm c}$. Polarized neutron scattering reveals that the single resonance is mostly isotropic in spin space, similar to overdoped NaFe$_{0.935}$Co$_{0.045}$As but different from optimal electron-, hole-, and isovalent-doped BaFe$_2$As$_2$ compounds, all featuring an additional prominent anisotropic component. Spin anisotropy in NaFe$_{0.9785}$Co$_{0.0215}$As is instead present at energies below the resonance, which becomes partially gapped below $T_{\rm c}$, similar to the situation in optimal-doped YBa$_2$Cu$_3$O$_{6.9}$. Our results indicate that  anisotropic spin fluctuations in NaFe$_{1-x}$Co$_x$As appear in the form of a resonance in the underdoped regime, become partially gapped below $T_{\rm c}$ near optimal doping and disappear in overdoped compounds.
   
\end{abstract}


\maketitle

\section{Introduction}

A common theme of unconventional superconductivity in iron pnictides is the interplay between superconductivity and magnetism, with stripe-type antiferromagnetic (AF) fluctuations potentially playing the role of a bosonic glue that binds Cooper pairs \cite{scalapino,dai,DSInosov_CRP}. An experimental determination of the evolution of spin fluctuations across the superconducting dome in iron pnictides is therefore important for the understanding of these fascinating materials. 

Parent compounds of iron pnictides such as $A$Fe$_2$As$_2$ ($A =$ Ca,Sr,Ba) and NaFeAs are antiferromagnetically ordered and the corresponding spin waves with bandwidths $\sim0.1$ eV have been carefully studied, and interpreted using effective Heisenberg models \cite{JZhao2009_NP,LWHarriger2011_PRB,CZhang2014_PRL} or itinerant electron models \cite{SODiallo2009_PRL,RAEwings2011_PRB}. Upon doping, magnetic order is gradually suppressed, and superconductivity is induced. Despite these changes, high-energy magnetic excitations resembling those in the parent compounds persist \cite{MSLiu2012_NP,KJZhou2013_NC,MWang2013_NC}. On the other hand, low-energy magnetic excitations in the normal state are significantly modified, with profile of the excitations in the $ab$-plane becoming more elongated along the transverse and longitudinal directions upon electron- and hole-doping, respectively \cite{JTPark2010_PRB,CZhang2011_SR}. Spin anisotropy gaps in the parent compounds are quickly suppressed \cite{LWHarriger2009_PRL}, replaced by overdamped and diffusive spin excitations \cite{GSTucker2014_PRB}. For superconducting samples, the most prominent change is the development of a spin resonance mode in the superconducting state, with intensity tracking the superconducting order parameter and also observed in other families of unconventional superconductors that exhibit strong magnetic fluctuations \cite{scalapino}. The resonance mode appears as a significant enhancement of magnetic fluctuations in the superconducting state relative to the normal state, present only at well-defined momentum and energy transfers. Appearance of the resonance is usually accompanied by a complete or partial gapping of magnetic spectral weight below the resonance mode, with the total magnetic spectral weight being conserved.

Energy of the resonance mode ($E_{\rm r}$) well inside the superconducting state has been proposed to universally scale with either the superconducting transition temperature $T_{\rm c}$ ($E_{\rm r}\approx$ 4-6 $k_{\rm B}T_{\rm c}$)  \cite{SDWilson,MWang2010_PRB} or the superconducting gap $2\Delta$ ($E_{\rm r}\approx0.64\times2\Delta$) \cite{GYu}, although some iron-based superconductors deviate from such scalings \cite{DSInosov2011_PRB,CZhang2016_PRB,QWang2016_PRL,CHLee2016_SR}. The energy of the resonance mode was also found to track the superconducting order parameter as a function of temperature in optimal-electron-doped BaFe$_{1.85}$Co$_{0.15}$As$_2$ \cite{DSInosov2010_NP}. However, energies of the resonance modes in YBa$_2$Cu$_3$O$_7$ \cite{HFFong1997_PRB}, CeCoIn$_5$ \cite{CStock2008_PRL}, FeTe$_{0.6}$Se$_{0.4}$ \cite{LWHarriger2012_PRB}, optimal-hole-doped Ba$_{0.67}$K$_{0.33}$Fe$_2$As$_2$ \cite{MWang2013_NC} and electron-overdoped NaFe$_{0.935}$Co$_{0.045}$As \cite{CZhang2013_PRB} were found to only slightly soften on approaching $T_{\rm c}$.
 
Recent angle-resolved photoemission spectroscopy (ARPES) measurements indicated spin-orbit coupling to be important for understanding the low-energy electronic structure of iron-based superconductors \cite{PDJohnson2015_PRL,Borisenko,ACharnukha}. Spin-orbit coupling also accounts for the ``\emph{XYZ}" spin anisotropy in parent compounds of iron pnictides \cite{NQureshi2012_PRB,CWang2013_PRX,YSong2013_PRB} and spin anisotropy in iron-based superconductors \cite{YSong2016_PRB,MMa2017_PRX}. Spin anisotropy is manifested in the resonance mode of superconducting iron pnictides up to optimal or slightly overdoped regime in electron- and hole-doped BaFe$_2$As$_2$ \cite{PSteffens2013_PRL,CZhang2013_PRB2,NQureshi2014_RPB}, leading to resonance modes that exhibit an anisotropic part in addition to an isotropic part in spin space. The resonance mode becomes fully isotropic well into the overdoped regime \cite{MSLiu2012_PRB}. With unpolarized neutron scattering the two parts are difficult to resolve, and it has been suggested the two parts may have different origins \cite{MWang2016_PRB}.

In underdoped NaFe$_{1-x}$Co$_x$As ($x=0.015$) superconductor, two resonance modes are resolved even with unpolarized neutron scattering \cite{CZhang2013_PRL}. Using polarized neutron scattering, it was further found the mode at lower energy is anisotropic in spin space while the one at higher energy is isotropic \cite{CZhang2014_PRB}. With increasing doping the low-energy mode gradually loses spectral weight, while the mode at higher energy is present across the superconducting dome \cite{CZhang2016_PRB}, however it is unclear how doping affects spin fluctuations in terms of spin anisotropy.

In this work, we use unpolarized and polarized neutron scattering to study the temperature and polarization dependence of spin fluctuations in nearly-optimal-doped NaFe$_{0.9785}$Co$_{0.0215}$As superconductor exhibiting weak AF order with an ordered moment $\approx0.02\mu_{\rm B}$ \cite{GTan2016_PRB}. A single resonance mode is observed, in contrast to underdoped NaFe$_{0.985}$Co$_{0.015}$As \cite{CZhang2013_PRL}. We find intensity of the resonance mode follows the superconducting order parameter, but energy of the mode $E_{\rm r}$ is almost temperature-independent, softening only slightly on approaching $T_{\rm c}$. This is different from the behavior in optimal-doped BaFe$_{1.85}$Co$_{0.15}$As$_2$, but similar to other unconventional superconductors \cite{HFFong1997_PRB,CStock2008_PRL,LWHarriger2012_PRB,MWang2013_NC,CZhang2013_PRB}. Polarized neutron scattering reveals the resonance to be mostly isotropic, different from electron-, hole- and isovalent-doped BaFe$_2$As$_2$ superconductors near optimal doping \cite{PSteffens2013_PRL,CZhang2013_PRB,DHu}, which also exhibit an additional anisotropic component. Significant spin anisotropy is instead present below the resonance mode, in the form of remnant spectral weight inside a partial spin gap induced by superconductivity, similar to optimal-doped YBa$_2$Cu$_{3}$O$_{6.9}$ \cite{NSHeadings2011_PRB}. Spin anisotropy at these energies persists up to $T\approx35$ K, and measurements of resistivity change under unixial stress indicates Curie-Weiss behavior down to a similar temperature. This finding confirms the link between deviation from Curie-Weiss behavior in nematic susceptibility and development of low-energy spin anisotropy, previously found in BaFe$_2$As$_2$-derived superconductors \cite{YSong2016_PRB}. Combined with previous results \cite{CZhang2014_PRB}, our work establishes a systematic understanding of how spin anisotropy evolves with doping in superconducting NaFe$_{1-x}$Co$_x$As.

\begin{figure}
\includegraphics[scale=0.5]{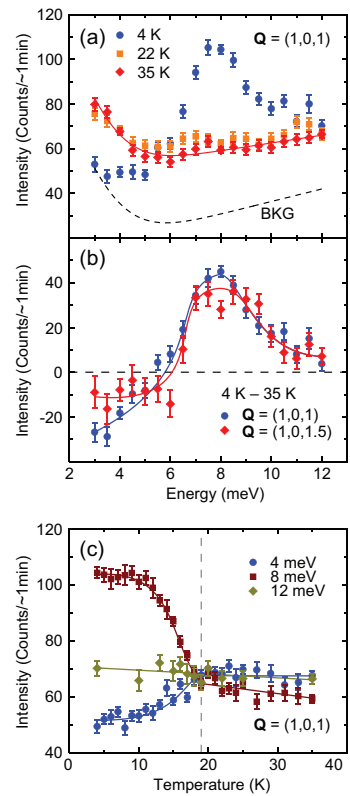} \protect\caption{ (Color online) (a) Constant-${\bf Q}$ scans at ${\bf Q}=(1,0,1)$ for $T=4$, 22 and 35 K. The solid line is an empirical fit to data at 35 K. The dashed line is a fit to the background (BKG) intensity. (b) Comparison of difference between 4 K and 35 K for constant-${\bf Q}$ scans at ${\bf Q}=(1,0,1)$ and $(1,0,1.5)$. The solid lines are guides to the eye. (c) Temperature dependence of spin fluctuations for $E = 4$, 8 and 12 meV at ${\bf Q}=(1,0,1)$. The solid lines are guides to the eye. Data in this figure are obtained on HB-3 using unpolarized neutron scattering.}
\end{figure}

\section{Experimental Details}
 
Single crystals of NaFe$_{0.9785}$Co$_{0.0215}$As were grown using the self-flux method \cite{NSprison2012_PRB} and were previously studied using elastic neutron scattering \cite{GTan2016_PRB} and time-of-flight neutron spectroscopy \cite{SVCarr2016_PRB}. Unpolarized inelastic neutron scattering measurements were carried out using the HB-3 thermal triple-axis spectrometer at the High Flux Isotope Reactor, Oak Ridge National Laboratory. Polarized inelastic neutron scattering measurements were carried out using the IN22 triple-axis spectrometer at Institut Laue-Langevin. Fixed $E_{\rm f}=14.7$ meV was used for both experiments. The experiment on HB-3 used a pyrolitic graphite monochromtor, analyzer, and filter after the
sample, the collimation used is 48$^\prime$-40$^\prime$-sample-40$^\prime$-120$^\prime$. The experiment on IN22 used a Heusler monochromator and analyzer, and utilizes the CRYOPAD for longitudinal polarization analysis. We adopt notation for the orthorhombic structural unit cell of NaFeAs ($a\approx b\approx5.56$ {\AA }, $c=6.95$ {\AA }), and aligned samples in the $[H,0,L]$ scattering plane to access excitations at ${\bf Q}=(1,0,L)$. For NaFe$_{1-x}$Co$_x$As displaying magnetic order, half-integer $L$-values correspond to AF zone centers and integer $L$-values correspond to AF zone boundaries along $c$-axis \cite{SLLi}. 

For polarized neutron scattering three neutron spin-flip (SF) cross sections $\sigma_{x}^{{\rm SF}}$, $\sigma_{y}^{{\rm SF}}$ and $\sigma_{z}^{{\rm SF}}$ were measured,
with the usual convention $x\parallel{\bf Q}$, $y\perp{\bf Q}$ in the scattering plane, and $z$ perpendicular to the scattering plane. Magnetic scattering polarized along $\alpha$-direction, $M_{\alpha}$ ($\alpha=y,z$), can be obtained from measured SF cross sections through $\sigma_{x}^{{\rm SF}}-\sigma_{y}^{{\rm SF}}\propto M_{y}$ and $\sigma_{x}^{{\rm SF}}-\sigma_{z}^{{\rm SF}}\propto M_{z}$  \cite{YSong2016_PRB}. By comparison, unpolarized neutron scattering does not separate these two quantities, and the measured cross sections contain both $M_y$ and $M_z$. With our experiment geometry, at ${\bf Q}=(1,0,L)$, $M_y$ is a combination of $M_a$ and $M_c$ whereas $M_z=M_b$. Spin-anisotropic magnetic fluctuations can be observed through differing $\sigma_{y}^{{\rm SF}}$ and $\sigma_{z}^{{\rm SF}}$ cross sections, and differing $M_y$ and $M_z$.

\section{Results}
\subsection{Temperature Dependence of Spin Fluctuations From Unpolarized Neutron Scattering}

Constant-${\bf Q}$ scans at ${\bf Q}=(1,0,1)$ for $T=4$ K (well below $T_{\rm c}$), 22 K (just above $T_{\rm c}$) and 35 K (just above $T_{\rm N}$) are compared in Fig. 1(a). Similar to slightly overdoped NaFe$_{0.935}$Co$_{0.045}$As, a single resonance mode forms at the expense of spectral weight at lower energies in the superconducting state \cite{CZhang2013_PRB}. The results at 22 K and 35 K are similar, consistent with the small ordered moment of $\approx0.02\mu_{\rm B}$ \cite{GTan2016_PRB} having little impact on magnetic fluctuations  below $T_{\rm N}$. In Fig. 1(b), we compare the difference of magnetic intensity between 4 K and 35 K for ${\bf Q}=(1,0,1)$ and ${\bf Q}=(1,0,1.5)$. Whereas the resonance mode shows little dependence on $L$, reduction of spectral weight below the resonance mode is more significant for integer $L$. Given that in the normal state, magnetic fluctuations at AF zone center (half-integer $L$) is at least as strong as fluctuations at AF zone boundary along $c$-axis (integer $L$), the smaller reduction of spectral weight at AF zone center implies significant remnant spectral weight below the resonance mode at AF zone center in the superconducting state. Temperature dependence of the scattering at ${\bf Q}=(1,0,1)$ is shown for several representative energies in Fig. 1(c). At the resonance energy $E=8$ meV, an order-parameter-like behavior is seen. Below the resonance, a clear reduction of spectral weight is observed.
Above the resonance energy, intensity of magnetic excitations does not respond to the onset of superconductivity. Such temperature dependence is similar to other iron pnictide superconductors \cite{DSInosov2010_NP,CZhang2013_PRL}.    

Constant-${\bf Q}$ scans at several temperatures below $T_{\rm c}\approx19$ K are shown in Fig. 2(a), subtracted by a fit to 35 K data to highlight the resonance mode and reduction of spectral weight at lower energies. The data in Fig. 2(a) are color-coded and interpolated, as shown in Fig. 2(b) for a direct visualization. To quantitatively characterize temperature dependence of the resonance mode, data points in Fig. 2(a) with $5\le E\le10$ meV are fit to Gaussian peaks to extract energy and intensity of the mode, with results shown in Fig. 2(c). Whereas intensity of the resonance can be reasonably described by the BCS order parameter with $T_{\rm c}\approx 19$ K \cite{KTerashima}, energy of the resonance only slightly softens from $E\approx8$ meV to $E\approx7$ meV.

\begin{figure}
	\includegraphics[scale=0.45]{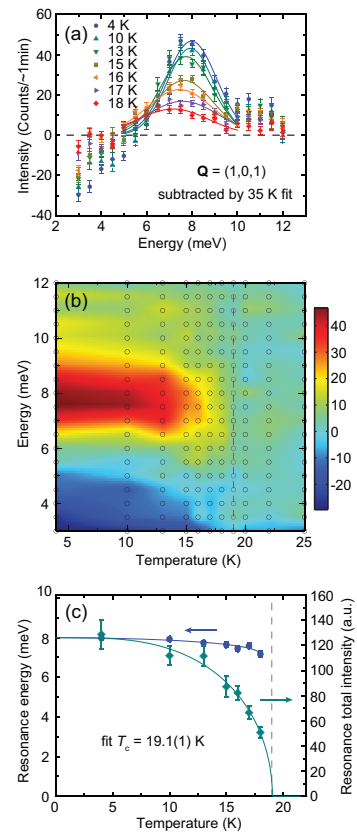} \protect\caption{(Color online) (a) Constant-${\bf Q}$ scans at ${\bf Q}=(1,0,1)$ for several temperatures below $T_{\rm c}$, subtracted by the empirical fit to 35 K data. The solid lines are fits to Gaussian peaks in the energy range  $5\le E\le10$ meV. (b) Color-coded and interpolated temperature dependence of low-energy magnetic fluctuations. The empirical fit to 35 K data has been subtracted. The circles correspond to points where measurements were taken. (c) Temperature dependence for the center and the area of the resonance mode, obtained from Gaussian fits in (a). The solid line for the resonance energy is a guide to the eye, and the solid line for the total area is a fit to the superconducting order parameter \cite{KTerashima}. The dashed vertical lines represent $T_{\rm c}\approx19$ K. Data in this figure are obtained on HB-3 using unpolarized neutron scattering.}
\end{figure}

An alternative way to extract temperature dependence for energy of the resonance mode is to directly examine the magnetic intensity in the superconducting state, without subtracting the normal state response, as shown in Fig. 3. Results in Fig. 3(a) can be phenomenologically modeled as damped harmonic oscillator responses, with measured intensity $I({\bf Q},E)\propto\frac{\chi^{\prime\prime}({\bf Q},E)}{1-\exp{(-\frac{E}{k_{\rm B}T})}}$ and $\chi^{\prime\prime}({\bf Q},E)\propto\frac{E_{0}^2\gamma E}{(E_0^2-E^2)^2+\gamma^2 E^2}$ \cite{GSTucker2014_PRB}. $E_0$ characterizes energy of the mode while $\gamma$ characterizes damping of the mode. The resulting $E_0$ and $\gamma$ for different temperatures are shown in Fig. 3(c) together with $E_{\rm max}$, the energy at which $\chi^{\prime\prime}({\bf Q},E)$ is maximized. As can be seen, both $E_0$ and $E_{\rm max}$ change only slightly with temperature. While $E_0$ depends weakly on temperature, $\gamma$ increases significantly with increasing temperature. Interestingly, the above analysis indicates appearance of the resonance mode in NaFe$_{0.9785}$Co$_{0.0215}$As can be interpreted as removal of damping from an existing mode, a scenario recently proposed for Ce$_{1-x}$Yb$_x$CoIn$_5$ \cite{YSong2016_NC115}. However, we note unlike Ce$_{1-x}$Yb$_{x}$CoIn$_5$, behaviors of the resonance mode in iron pnictides are also consistent with the spin-exciton scenario \cite{CZhang2013_PRB,MGKim2013_PRL}.

From analysis presented in Fig. 2 and Fig. 3, we demonstrate that energy of the resonance mode in NaFe$_{0.9785}$Co$_{0.0215}$As depends weakly on temperature, this conclusion holds whether we analyze our data by subtracting the normal state response (Fig. 2) or not (Fig. 3). 

\begin{figure}
	\includegraphics[scale=0.45]{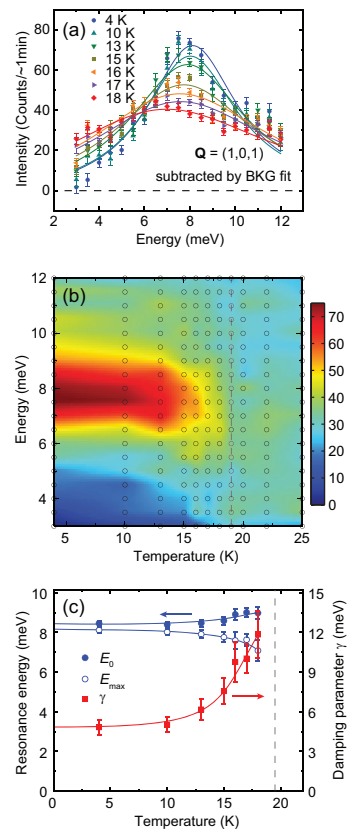} \protect\caption{(Color online) (a) Constant-${\bf Q}$ scans at ${\bf Q}=(1,0,1)$ for several temperatures below $T_{\rm c}$, subtracted by the fit to background (BKG) intensity. The solid lines are fits to damped harmonic oscillator responses in the energy range  $3\le E\le12$ meV. (b) Color-coded and interpolated temperature dependence of low-energy magnetic fluctuations. The fit to background has been subtracted. The circles correspond to points where measurements were taken. (c) Temperature dependence for $E_0$, $E_{\rm max}$ and $\gamma$ from damped harmonic oscillator fits in (a). The solid lines are guides to the eye. The dashed vertical lines represent $T_{\rm c}\approx19$ K. Data in this figure are obtained on HB-3 using unpolarized neutron scattering.}
\end{figure}

\subsection{Polarization of Spin Fluctuations From Polarized Neutron Scattering}

Constant-${\bf Q}$ scans of the three SF cross sections $\sigma_{x}^{{\rm SF}}$, $\sigma_{y}^{{\rm SF}}$
and $\sigma_{z}^{{\rm SF}}$ at ${\bf Q}=(1,0,0.5)$ were measured well below ($T=2$ K) and just above $T_{\rm c}$ ($T=21$ K), and are shown in Figs. 4(a) and (b). Magnetic fluctuations inside the superconducting state are clearly modified from their normal state counterpart, displaying both the resonance mode and a superconductivity-induced spin gap, in agreement with unpolarized neutron scattering results in Fig. 1. Despite such changes, spin anisotropy as indicated by differing  $\sigma_{y}^{{\rm SF}}$ and $\sigma_{z}^{{\rm SF}}$ is observed below a similar energy ($E<7$ meV) for both temperatures. The differences, $\sigma_{x}^{{\rm SF}}-\sigma_{y}^{{\rm SF}}\propto M_y$ and $\sigma_{x}^{{\rm SF}}-\sigma_{z}^{{\rm SF}}\propto M_z$, are shown in Figs. 4(c) and (d) for the two temperatures. In the superconducting state [Fig. 4(c)], while $M_z$ is gapped for $E\lesssim$ 5 meV, significant spectral weight in $M_y$ remains. The remnant spectral weight in $M_y$ therefore accounts for the partial gapping of spectral weight at half-integer $L$ seen in unpolarized neutron scattering [Fig. 1(b)]. In the normal state [Fig. 4(d)], spin anisotropy is also observed, similar to electron- and hole-doped BaFe$_2$As$_2$ \cite{PSteffens2013_PRL,HQLuo2013,CZhang2013_PRB2}, but different from isovalent-doped BaFe$_{2}$As$_{1.4}$P$_{0.6}$ \cite{DHu}. 

\begin{figure}
	\includegraphics[scale=0.45]{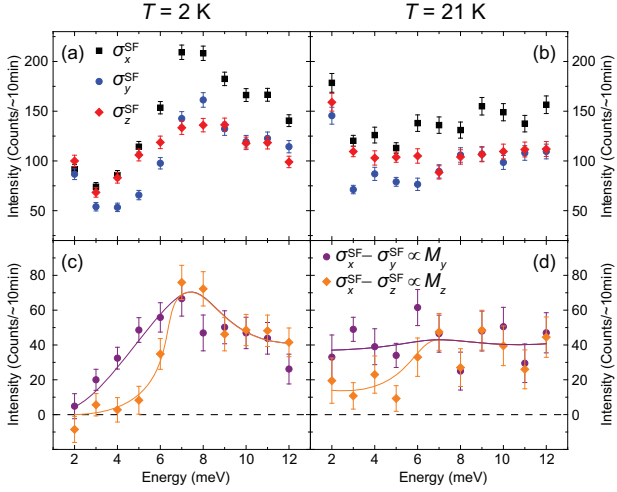} \protect\caption{ (Color online) Constant-${\bf Q}$ scans of $\sigma_{x}^{{\rm SF}}$, $\sigma_{y}^{{\rm SF}}$
		and $\sigma_{z}^{{\rm SF}}$ at ${\bf Q}=(1,0,0.5)$ (a) well below $T_{\rm c}$ ($T=2$ K) and (b) just above $T_{\rm c}$ ($T=21$ K). The differences $\sigma_{x}^{{\rm SF}}-\sigma_{y}^{{\rm SF}}$
		and $\sigma_{x}^{{\rm SF}}-\sigma_{z}^{{\rm SF}}$, which are respectively
		proportional to $M_{y}$ and $M_{z}$, are correspondingly shown in (c) and (d). The solid lines are guides to the eye. Data in this figure are obtained on IN22 using polarized neutron scattering.}
\end{figure}

In BaFe$_2$As$_2$-derived superconductors near optimal doping, a prominent anisotropic contribution to the resonance mode is also observed \cite{PSteffens2013_PRL,CZhang2013_PRB2,YSong2016_PRB,DHu}. Comparing the results in Figs. 4(c) and (d), 
while anisotropic fluctuations at $E=6$ meV may be slightly enhanced in the superconducting state
of nearly optimal-doped NaFe$_{0.9785}$Co$_{0.0215}$As, most of the anisotropic magnetic fluctuations reside below the resonance mode, becoming partially gapped inside the superconducting state. This conclusion is corroborated by temperature dependence of spin anisotropy measured for $E=3$ and 5 meV, shown in Fig. 5. For $T\gtrsim35$ K, both energies display $\sigma_{y}^{{\rm SF}}\approx\sigma_{z}^{{\rm SF}}$, indicating isotropic magnetic fluctuations [Figs. 5(a) and (b)]. The evolution of $M_y$ and $M_z$ as a function of temperature obtained from the differences $\sigma_{x}^{{\rm SF}}-\sigma_{y}^{{\rm SF}}$ and $\sigma_{x}^{{\rm SF}}-\sigma_{z}^{{\rm SF}}$ are shown in Figs. 5(c) and (d). For $E=3$ meV, $M_y$ increases on approaching $T_{\rm c}$ from above and is suppressed below $T_{\rm c}$ whereas $M_z$ displays little temperature dependence above $T_{\rm c}$ and is also suppressed below $T_{\rm c}$. For $E=5$ meV, $M_y$ displays little temperature dependence whereas $M_z$ behaves similarly to $E=3$ meV. At both energies, the magnetic fluctuations display clear spin anisotropy for $T\lesssim35$ K but no enhancement is observed below $T_{\rm c}$ in either $M_y$ or $M_{z}$. This coupled with the observation that spin anisotropy is only present for $E<7$ meV [Figs. 4(a) and (b)] suggests potential anisotropic fluctuations that become enhanced in the superconducting state could only exist for $5<E<7$ meV, although no such indication can be seen in the constant-${\bf Q}$ scans [Figs. 4(c) and (d)]. 
 
\begin{figure}
\includegraphics[scale=0.45]{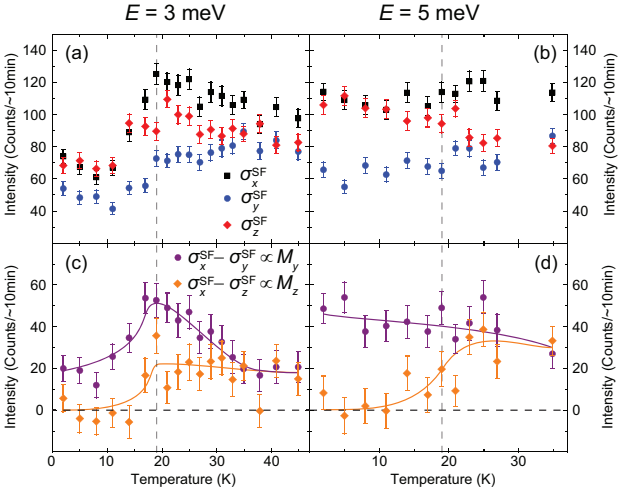} \protect\caption{(Color online) Temperature dependence of $\sigma_{x}^{{\rm SF}}$, $\sigma_{y}^{{\rm SF}}$
	and $\sigma_{z}^{{\rm SF}}$ at ${\bf Q}=(1,0,0.5)$ for (a) $E=3$ meV and (b) $E=5$ meV. The differences $\sigma_{x}^{{\rm SF}}-\sigma_{y}^{{\rm SF}}$
	and $\sigma_{x}^{{\rm SF}}-\sigma_{z}^{{\rm SF}}$, which are respectively
	proportional to $M_{y}$ and $M_{z}$ are correspondingly shown in (c) and (d). The solid lines are guides to the eye. The vertical dashed lines represent $T_{\rm c}$. Data in this figure are obtained on IN22 using polarized neutron scattering.}
\end{figure}

\subsection{Resistivity Change Under Uniaxial Pressure}

Previously, it was found the resistivity change under uniaxial strain (elastoresistance) \cite{Kuo2016} or stress \cite{ZYLiu2016,YHGu2017}, which acts as a proxy for the nematic susceptibility, displays Curie-Weiss (CW) temperature dependence in many iron-based superconductors. However, for electron- and hole-doped BaFe$_2$As$_2$ superconductors, deviations from CW behavior were found at temperatures above $T_{\rm c}$, although CW behavior was found down to $T_{\rm c}$ in isovalent-doped BaFe$_{2}$As$_{1.4}$P$_{0.6}$ \cite{Kuo2016}. It was noted that the temperatures at which nematic susceptibility deviate from CW behavior correspond to temperatures at which spin anisotropy onset in these systems, suggesting anisotropic magnetic fluctuations may be responsible for the deviation from CW behavior in nematic susceptibility \cite{YSong2016_PRB}.  

\begin{figure}
	\includegraphics[scale=0.45]{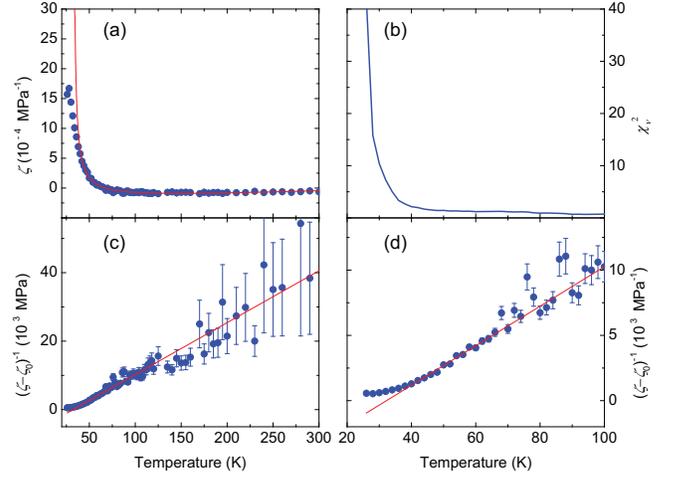} \protect\caption{(Color online) (a) Resistivity change under uniaxial stress $\zeta$ along (100) direction for the orthorhombic unit cell [(110) direction for the tetragonal unit cell]. The solid red line is a CW fit for $T\geq40$ K. (b) Reduced chi-squared from CW fits of data in (a), by setting the fitting ranges to be from different temperatures to 300 K. The same standard deviation is used for all measured data points to obtain reduced chi-squared, and is estimated from the standard deviation of data with $T>100$ K. (c) $(\zeta-\zeta_0)^{-1}$ for the fit shown in (a), the error bars for $\zeta$ is estimated from standard deviation of data with $T>100$ K, and the error bars for $(\zeta-\zeta_0)^{-1}$ is obtained through propagation of error. (d) Zoom-in of results in (c).}
\end{figure}

Having established anisotropic magnetic excitations onset at $T\approx35$ K in NaFe$_{0.9785}$Co$_{0.0215}$As (Fig. 5), it would be interesting to check if CW behavior in nematic susceptibility holds down to a similar temperature. To this end, we measured resistivity change under uniaxial stress $\zeta$ using the device described previously \cite{ZYLiu2016}, with stress applied along (100) direction of the orthorhombic unit cell [(110) direction of the tetragonal unit cell], and the result is shown in Fig. 6(a). The data is fit to the CW form $\zeta=\zeta_0+\frac{A}{T-T_{\rm CW}}$. To account for a weak upturn observed for $T\gtrsim200$ K, $\zeta_0$ is allowed to have a weak linear dependence on temperature, rather than being fully temperature-independent. 

A reasonable fit is obtained by fitting the data from 40 K to 300 K [solid red lines in Fig. 6(a)]. From the data and fit in Fig. 6(a), $(\zeta-\zeta_0)^{-1}=\frac{T-T_{\rm CW}}{A}$ is shown in Figs. 6(c) and 6(d), with Fig. 6(d) zoomed in to focus on data with $T<100$ K. Linear behavior in $(\zeta-\zeta_0)^{-1}$ is seen from 300 K down to $T\approx40$ K [Fig. 6(d)], and clear deviation from linear behavior is seen at lower temperature. From the fit in Fig. 6(a) we obtain $T_{\rm CW}\approx31$ K and $A^{-1}\approx135$ MPa/K. The value of $T_{\rm CW}$ in our sample is between $T_{\rm CW}$ of NaFeAs and NaFe$_{0.986}$Ni$_{0.015}$As, and the value of $A^{-1}$ is reasonably close to those reported in NaFe$_{1-x}$Ni$_x$As \cite{YHGu2017}, after adjusting for a Fermi surface factor $\kappa\approx11$ resulting in $A_{n}^{-1}\approx12$ MPa/K \cite{YHGu2017}. However, we found that both $T_{\rm CW}$ and $A^{-1}$ depend on the fitting range we use. Fitting the data from 30 K to 300 K we obtain $T_{\rm CW}\approx21$ K and $A^{-1}\approx60$ MPa/K, while fitting the data from 50 K to 300 K we obtain $T_{\rm CW}\approx38$ K and $A^{-1}\approx234$ MPa/K.

Goodness of fit strongly depends on the chosen fitting range, as can be seen in reduced chi-squared ($\chi^{2}_{\nu}$) obtained by fitting starting from different temperatures to 300 K, as shown in Fig. 6(b). While $\chi^{2}_{\nu}$ changes only modestly when fitting starts from temperatures $T\gtrsim40$ K, it increases dramatically when the fitting starts from lower temperatures. This indicates $\zeta$ deviates from CW behavior below $T\approx40$ K, close to $T\approx35$ K below which spin anisotropy develops. The persistence of CW behavior down to a similar temperature is also observed in nearly-optimal-doped NaFe$_{0.985}$Ni$_{0.015}$As \cite{YHGu2017}. These results indicate compared to electron-doped BaFe$_2$As$_2$ \cite{Kuo2016}, CW behavior in nematic susceptibility persists to lower temperatures in electron-doped NaFeAs, and confirms the link between development of anisotropic magnetic fluctuations and the nematic susceptibility deviating from CW behavior \cite{YSong2016_PRB}.   

\section{Discussion and Conclusion}

\begin{figure}
	\includegraphics[scale=0.55]{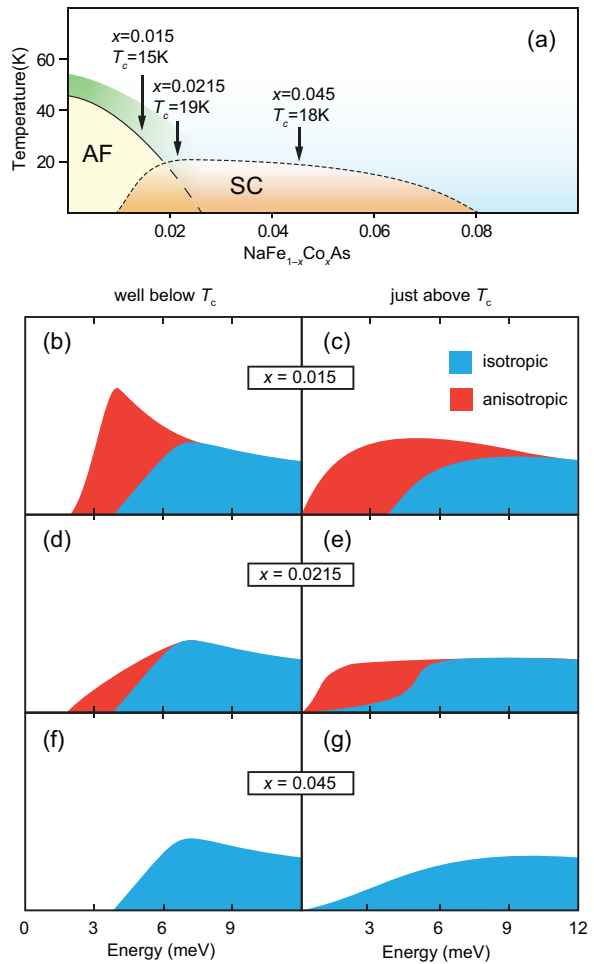} \protect\caption{(Color online) (a) Schematic phase diagram of NaFe$_{1-x}$Co$_x$As \cite{GTan2013}. The three concentrations for which polarized neutron scattering have been carried out are marked by arrows. Sketches of isotropic and anisotropic magnetic fluctuations well below $T_{\rm c}$ and just above $T_{\rm c}$ for $x=0.015$ are respectively shown in (b) and (c). Similar sketches for $x=0.0215$ are shown in (d) and (e), and for $x=0.045$ in (f) and (g).}
\end{figure}

Our polarized neutron scattering results reveal that anisotropic spin fluctuations in nearly-optimal-doped NaFe$_{0.9785}$Co$_{0.0215}$As become partially suppressed inside the superconducting state, lacking the prominent anisotropic resonance mode that is enhanced with the onset of superconductivity, seen in BaFe$_2$As$_2$-derived superconductors nearly optimal doping. Instead, the presence of anisotropic fluctuations that exist below an isotropic resonance mode is similar to what is observed in optimal-doped YBa$_2$Cu$_3$O$_{6.9}$ \cite{NSHeadings2011_PRB}. Previously, it was found that of the two resonance modes in underdoped NaFe$_{0.985}$Co$_{0.015}$As, the mode at lower energy is anisotropic while the one at higher energy is isotropic \cite{CZhang2014_PRB}. With Co-doping, the resonance mode at lower energy becomes suppressed near optimal doping, while the mode at higher energy persists across the superconducting dome \cite{CZhang2016_PRB}. These findings are consistent with our present conclusion that no prominent anisotropic resonance mode is present in nearly-optimal-doped NaFe$_{0.9785}$Co$_{0.0215}$As.

Combined with previous results \cite{CZhang2014_PRB}, our present work allows for a systematic understanding of how anisotropic spin dynamics evolve in NaFe$_{1-x}$Co$_x$As, as sketched in Fig. 7. Three samples have been so far studied, representative of underdoped ($x=0.015$), optimal-doped ($x=0.0215$) and overdoped ($x=0.045$) regions of the phase diagram [Fig. 7(a)]. As can be seen, with increasing doping, anisotropic fluctuations are gradually suppressed, evolving from a resonance mode in the underdoped regime [Fig. 7(b)] to remnant spectral inside a superconductivity-induced partial spin gap near optimal doping [Fig. 7(d)], and disappearing in the overdoped regime [Fig. 7(f)]. 

It is noteworthy that anisotropic fluctuations typically appear above $T_{\rm c}$ [Figs. 7(c) and (e)], and develop into an anisotropic resonance mode or become partially gapped inside the superconducting state \cite{PSteffens2013_PRL,CZhang2013_PRB2,HQLuo2013,CZhang2014_PRB,YSong2016_PRB,MMa2017_PRX}. BaFe$_2$As$_{1.4}$P$_{0.6}$ appears to be an exception, with anisotropic spin fluctuations only present in the superconducting state \cite{DHu}.  

Anisotropic fluctuations at the stripe-type AF ordering wave vector in iron pnictides and chalcogenides reported so far can be viewed to fall within situations of Fig. 7. In one extreme, there is FeSe with the resonance entirely anisotropic (although it is unclear at what energy fluctuations become isotropic), and in the other extreme the resonance is fully isotropic as found in overdoped iron pnictides \cite{MSLiu2012_PRB,CZhang2014_PRB}. In most systems reported so far, it is observed $M_y\geq M_z$, namely $c$- or $a$-axis (in-plane longitudinal direction at the stripe vector) polarized excitations are at least as intense as $b$-axis (in-plane transverse direction at the stripe vector) polarized excitations. However, recently it was reported that in underdoped Ba(Fe$_{0.955}$Co$_{0.045}$)$_2$As$_2$ with coexisting magnetic order and superconductivity, the resonance mode has no isotropic component \cite{Wasser2017_SR}, different from the behaviors depicted in Fig. 7. A possible reason is that the ordered moment in Ba(Fe$_{0.955}$Co$_{0.045}$)$_2$As$_2$ is $\approx0.2\mu_{\rm B}$/Fe, much larger than $\approx0.03\mu_{\rm B}$/Fe seen in underdoped NaFe$_{0.985}$Co$_{0.015}$As \cite{GTan2016_PRB}, and therefore features stronger interplay between magnetic order and the resonance mode.  

While we have linked the development of spin anisotropy with deviation from CW behavior in nematic susceptibility in iron pnictides near optimal doping, diverging longitudinal fluctuations in BaFe$_2$As$_2$ just above $T_{\rm N}$ \cite{YLi2017} does not have the same effect on nematic susceptibility, with CW nematic susceptibility persisting down to $T_{\rm N}$ \cite{JHChu2012}. A possible cause for this difference is that whereas divergent longitudinal fluctuations in BaFe$_2$As$_2$ just above $T_{\rm N}$ have $M_a>M_b\approx M_c$ \cite{YLi2017}, anisotropic fluctuations in the normal state of BaFe$_2$As$_2$-derived superconductors near optimal doping exhibit $M_a\approx M_c>M_b$ \cite{HQLuo2013,YSong2016_PRB}. The differing character of polarization may account for the different effects on the nematic susceptibility.

In conclusion, we have studied the temperature and polarization dependence of magnetic fluctuations in nearly-optimal-doped NaFe$_{0.9785}$Co$_{0.0215}$As. While intensity of the resonance mode tracks the superconducting order parameter, energy of the mode only slightly softens approaching $T_{\rm c}$. Anisotropic fluctuations in NaFe$_{0.9785}$Co$_{0.0215}$As are present for $E<7$ meV and $T\lesssim35$ K, and anisotropic fluctuations mostly become partially gapped inside the superconducting state, lacking the anisotropic resonance mode seen in BaFe$_2$As$_2$ superconductors near optimal doping. Nonetheless, behavior of anisotropic fluctuations in iron-based superconductors can be viewed to qualitatively reside somewhere along a continuous evolution, as exemplified by the evolution of anisotropic fluctuations in NaFe$_{1-x}$Co$_x$As. However, the behavior of particular compounds appears to be material-specific, likely resulting from the interplay of superconducting gap energies and spin-orbit coupling.  

\section{Acknowledgments} We thank Zhuang Xu and Huiqian Luo for assistance in resistivity change under uniaxial stress measurements.
The single crystal growth and neutron
scattering work at Rice is supported by the U.S. DOE, BES
under Contract No. DE-SC0012311 (P.D.). The materials work at Rice is also supported by the Robert
A. Welch Foundation Grant Nos. C-1839 (P.D.). This research used resources at the High Flux Isotope Reactor, a DOE Office of Science User Facility operated by the Oak Ridge National Laboratory.

\end{document}